\documentclass[12pt]{article}
\usepackage{graphicx} 

\newcommand{\bq}{\begin{eqnarray}}
\newcommand{\eq}{\end{eqnarray}}

\begin{document}

\title{Small Volatility Approximation and Multi-Factor HJM Models.}
\author{V.M. Belyaev \\
US Bancorp, Minneapolis, MN, USA}


\maketitle
\begin{abstract}
Here we demonstrate how we can use Small Volatility Approximation in calibration
 of  Multi-Factor HJM model with deterministic correlations, factor volatilities and mean reversals.
It is noticed that quality of this calibration is very good and it does not depend on number of factors.
\end{abstract}

\section{Introduction.}
HJM model \cite{HJM} is a Normal Volatility Model which can be used to model forward rate dynamic.
Multi-Factor HJM model \cite{multi} can significantly improve calibration accuracy. Note, that Multi-Factor model have the following input deterministic input parameters:
factor volatilities, factor correlation and factor mean-reversion speeds/autocorrelations.
It means that forward volatilities are normal distributed and can not reproduce implied volatility smiles.

From other side there is a Small Volatility Approximation. This approximation can be used to calibrate all ATM swaptions.
Calibration process is very fast and accurate. 

Here we demonstrate how we can apply Small Volatility Approximation in Multi-Factor HJM model calibration.

\section{Small Volatility Approximation.}

It was found \cite{Belyaev} that Small Volatility Approximation has very good quality in calculation of ATM swaption prices.
This approximation can be used  to calibrate all available ATM swaptrions. Calibration procedure is fast and very accurate.

Single factor HJM model  is characterized by the following dynamics:
\bq
df(t,T)=\alpha(t,T)dt+\sigma(t,T)dW(t);
\eq
where $f(t,T)$ represent a forward rate:
\bq
B(t,T)=e^{-\int_t^Tf(t,\tau)d\tau};
\eq
$B(t,T)$ denotes a zero coupon risk-free  bond; $\sigma(t,T)$ is a deterministic normal volatility; $dW(t,T)$ represents a Brownian  motion; and
\bq
\alpha(t,T)=\sigma(t,T)\int_t^T\sigma(t,\tau)d\tau;
\eq
is a drift. 

This drift is chosen to satisfy the martingale condition on bond prices
\bq
B(0,T)=\left<
e^{-\int_0^tr(\tau)d\tau}B(t,T)
\right> ;\;\;\forall t\in[0,T].
\eq

In Small Volatility Approximations distribution of discounted bond prices at time $T$ is:
\bq
& & e^{-\int_0^Tr(\tau)d\tau}B(T,T_1)=
\nonumber
\\
& & =B(0,T_1)e^{-\int_0^Td\tau\int_\tau^{T_1}\alpha(\tau,t)dt-\int_0^TdW(\tau)\int_\tau^{T_1}\sigma(\tau,t)dt}=
\nonumber
\\
& & = B(0,T_1)\left(
1-\int_0^TdW(\tau)\int_\tau^{T_1}\sigma(\tau,t)dt+o(\sigma)
\right).
\eq
Within this approximation we can determine  swap price distributions.
Distribution of the SOFR swap  present value   is given by:
\bq
PV(T)=e^{-\int_0^Tr(t)dt}\sum_{n=1}^NB(T,T_n)\left(r_s+1-\frac{B(T,T_{n-1})}{B(T,T_n)}\right)=
\nonumber
\\
=e^{-\int_0^Tr(t)dt}\left(r_s\sum_{n=1}^NB(T,T_n)-
B(T,T)+B(T,T_N)
\right)\simeq
\nonumber
\\
\simeq (r_s-r_{ATM})\sum_{n=1}^NB(0,T_n)+ \Sigma(T,N)\xi\sqrt{T};
\label{PV}
\eq
where $T_n$ represent times to the payments; $r_s$ and $r_{ATM}=\frac{B(0,T)-B(0,T_N)}{\sum_{n=1}^NB(0,T_n)}$ denote  swap  rate  and ATM rate;
 $\xi$ is a standard normal distributed stochastic variable
$$
<\xi>=0;\;\;\;<\xi^2>=1;
$$
\bq
& & \Sigma^2(T,N)T=\int_0^Tv^2(t,N)dt;
\nonumber
\\
& & v(t,N)=r_s\sum_{n=1}^NB(0,T_n)\int_t^{T_n}\sigma(t,\tau)d\tau-
\nonumber
\\
& & -  B(0,T)\int_t^T\sigma(t,\tau)d\tau+B(0,T_N)\int_t^{T_N}\sigma(t,\tau)d\tau.
\label{sva}
\eq

Using formulas (\ref{sva}) we can calculate    swapton prices. It means that we can use them to determine forward bond volatilities from swaption prices.

\section{Calibration in Small Volatility Approximation.}

Let us consider a grid with 3-month time steps. The first swaption has a tenor of 1 and expires in 3 months. According to equation (\ref{sva}), we have:
\bq
& & v(dt,1)=r_s B(0,5dt)\sum_{k=0}^{4}(k+1)\sigma(0,k ) dt -
\nonumber
\\
& & -B(0,dt)\sigma(0,0)dt+
\nonumber
\\
& & +B(0,5dt)\sum_{k=0}^4(k+1)\sigma(0,k)dt;
\label{1}
\eq
where $dt=0.25$.

Assuming that all unknown volatilities are equal in equation (\ref{1})
\bq
\sigma(0,k)=\sigma(0,0);\;\forall k< 5;
\label{5}
\eq
 we can calculate volatilities in equation (\ref{sva}).

Next, we can extend this calculation to other expirations and tenors, using the previously defined volatility values. For each available tenor and time to expiration  $T_e$ we obtain the following equation:
\bq
D^2(T_e,tenor)\Sigma^2(T_e,tenor)=A\sigma^2+2B\sigma+C;
\label{eq}
\eq
where 
$A, B, C$ are factors which can be determine by using bond  prices and already calculated volatilities; $\sigma$ is an unknown forward volatility. 
In most cases, all parameters are non-negative, so we have
\bq
\sigma=
\frac{1}{A}\left(
-B+
\sqrt{B^2-
A(C-D^2(T_e,tenor)\Sigma^2(T_e,tenor)T_e)
}
\right).
\label{sigma}
\eq

Using this procedure, we can determine all forward bond volatilities.

 Applying this procedure we obtained good calibrated ATM swaption prices (June 3, 2025) depicted on Figure.1. We can see a small difference between model and input prices are observed only for long term swaptions with tenor 30.
 Note, that present values of swaptions in calibrated model are calculated exactly as:
 \bq
& &  PV(T_e=Ndt,tenor)=
 \nonumber
 \\
& &  =\left<
e^{-\int_0^{T_e}(t)dt}\left(r_s\sum_{n=1}^NB(T_e,T_n)-
B(T_e,T_e)+B(T_e,T_N)\right)
 \right>=
 \nonumber
 \\
 & & =\left<
 e^{-\sum_{n=0}^{N-1}f(ndt,ndt)dt-\sum_{m=N}^{M-1}f(Ndt,mdt)dt}
 \right>
 \eq

\section{Multi-Factor HJM Model.}

$N$-factor HJM model has the following dynamics:
\bq
df(t,T)=\alpha(t,T)dt+\sum_{n=1}^N\sigma_n(t,T)dW_n(t);
\label{HJM}
\eq
where $f(t,T)$ is a forward rate; $\sigma_n(t,T)$ is a deterministic forward volatility; $dW_n(t)$ is a Brownian motion; an $\alpha(t,T)$ is deterministic drift.

This drift can be determined from discounting bond prices:
\bq
& & e^{-\int_0^tr(\tau)d\tau}B(t,T)=B(0,T)e^{-\int_0^td\tau\int_\tau^T\alpha(\tau,t_1)dt_1-\sum_{n=1}^N\int_0^\tau dW_n(\tau)\int_\tau^T\sigma_n(\tau,t_1)dt_1};
\nonumber
\\
& & dW_n=\xi_n\sqrt{dt};\;\;\;\left<dW_m(t)dW_n(t)\right>=\rho_{mn}dt;
\label{B}
\eq
where $\xi_n$ is a normal distributed  stochastic variable;
\bq
<\xi_n>=0;\;\;\; <\xi_m\xi_n>=\rho_{mn};
\eq
 $\rho_{mn}$ is a correlation.

Eq.(\ref{B}) means that we have the following condition:
\bq
\left<e^{-dt\int_t^T\alpha(t,t_1)dt_1-\sum_{n=1}^N\xi_n \sqrt{dt}\int_t^T\sigma_n(t,t_1)dt_1
}\right>=1; \;\;\; \forall \;0<t\leq T.
\label{mar}
\eq

Using martingale condition   (\ref{mar}) we can calculate the drift for any N-Factor models.

Multi-factor HJM models can be used to calibrate ATM interest rate derivatives. Dynamics of this model has the following form:
\bq
f(t,T)=f(0,T)+\sum_{n=1}^N\int_0^t\alpha_n(\tau,T)d\tau+\sum_{n=1}^N\int_0^t\sigma_n(\tau,T)dW_n(\tau);
\eq
where $dW_n(\tau)$ are uncorrelated Brownian motions and
\bq
\alpha_n(\tau,T)=\sigma_n(\tau,T)\int_s^T\sigma_n(s,u)du.
\eq
$\alpha(\tau,T)$ is chosen to satisfy the martingale condition:
\bq
B(0,T)=\left<
e^{-\int_0^tr(\tau)d\tau}B(t,T)
\right>;\;\;\; \forall\; 0\leq t\leq T.
\eq

In this type of model we assume that volatilities are deterministic functions. It means that rate distributions are also normal and no volatility smile can be
generated.

This model can be calibrated to ATM swaptions by assuming that Brownian motions are
\bq
dW_m(t)=\sum_{n=1}^N w_{m,n}(t)d\tilde W_n(t);
\eq
where
$W_{m,n}(t)$ are correlated Brownian motions; and
\bq
\sigma_{m}(t,T)dW_m(t)=\sigma_m(t,0)e^{-\int_0^t\kappa_m(\tau)d\tau}.
\eq

Small Volatility Approximation we can determine all ATM swaption volatilities. So to calibrate this model, we need to select correlation and relative weights of
forward volatilities.

For example, let us consider two-factor volatility model and assume that 
\bq
\sigma_{1,2}(t,T)dW_{1,2}(t)=e^{-\kappa_{1,2}(t-T)}dW_{1,2}(t);
\eq

Using Small Volatility Approximation we have for time to expiration 0.25 and tenor 1:
\bq
PV(0.25,1)=v(0.1)\left(B(0,0.25)\xi_1+B(0,0.5)\xi_2+B(0,0.75)\xi_3+\right.
\nonumber
\\
\left. +B(0,1)\xi_4+B(0,1.25)\xi_5
\right)\sqrt{dt};
\eq
where
\bq
\xi_n\sqrt{dt}=\sigma_1((0,0)e^{-\kappa_1ndt}dW_1+\sigma_2((0,0)e^{-\kappa_2ndt}dW_2.
\eq
It is clear that because all model parameters are deterministic then distribution of forward volatiity will be normal and effectively we can adjust $v(0,1)$ to reproduce swaption price.
The same procedure will give us the same result for all expirations and tenors.

This procedure will give us the same quality of calibrated model as a single factor HJM model.
We need to have additional assumptions to build multi-factor HJM model. 
It can be taken by some Multi-Factor calibration procedure \cite{peter}.

\section{Conclusion.}

It is demonstrated how to increase quality of Multi-Factor model calibration by using Small Volatility Approximation.

\section{Disclaimer.}
The opinions expressed in this article are the author's own and they may be
different from the views of U.S. Bancorp.

\section{Figures}

\begin{figure}[h]
	\begin{minipage}{.5\textwidth}
		\includegraphics[width=50mm]{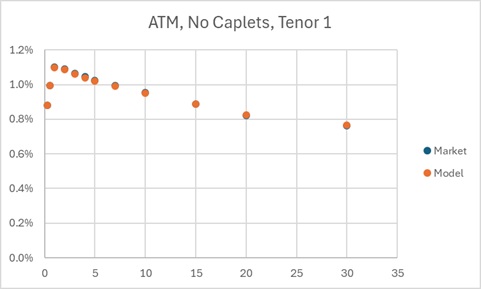}
	\end{minipage}
	\begin{minipage}{.5\textwidth}
		\includegraphics[width=50mm]{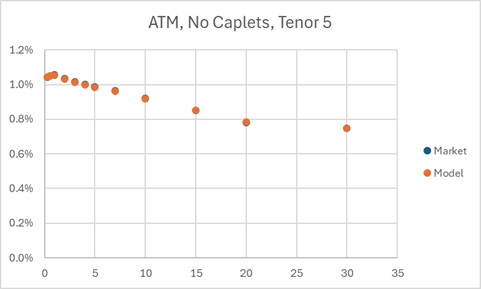}
	\end{minipage}
\end{figure}

\begin{figure}[h]
	\begin{minipage}{.5\textwidth}
		\includegraphics[width=50mm]{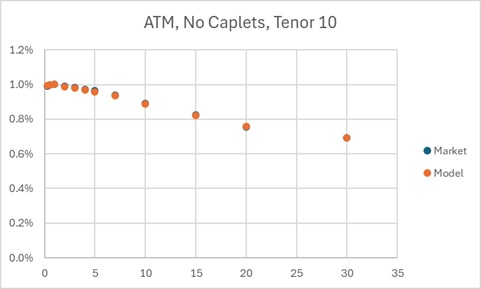}
	\end{minipage}
	\begin{minipage}{.5\textwidth}
		\includegraphics[width=50mm]{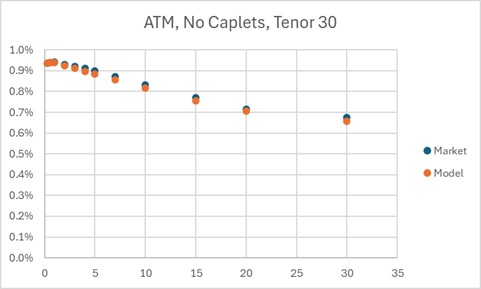}
	\end{minipage}
	\caption{ATM Market and Model Volatilities.}
	\label{ATM}
\end{figure}

\end{document}